\documentclass[acmsmall,screen]{acmart}

\usepackage{algorithm2e}
\usepackage{enumitem}
\usepackage{array}
\usepackage{wrapfig}
\usepackage{amsmath,amsfonts}
\usepackage[noend]{algpseudocode}
\usepackage{graphicx}
\usepackage{textcomp}
\usepackage{float}
\usepackage{listings}
\usepackage{xspace}
\usepackage{multirow}
\usepackage{amsthm}

\usepackage{balance}
\usepackage{algpseudocode}
\usepackage{colortbl}

\usepackage[skins]{tcolorbox}
\usepackage{xcolor,pifont}
\usepackage{multicol}
\newcommand*\colourcheck[1]{%
	\expandafter\newcommand\csname #1check\endcsname{\textcolor{#1}{\ding{52}}}%
}
\colourcheck{blue}
\colourcheck{green}
\colourcheck{red}
\newtcolorbox{boxB}[2][]{%
  enhanced,colback=white,colframe=black,coltitle=black,
  sharp corners,
  toprule=1.0pt,
  rightrule=0.3pt,
  leftrule=0pt,
  bottomrule=0pt,
  fonttitle=\itshape\scshape\large,
  left=0pt,right=5pt,top=5pt,bottom=3pt,
  attach boxed title to top right={yshift=-0.3\baselineskip-0.4pt,xshift=-5mm},
  boxed title style={tile,size=minimal,left=0.2mm,right=0.5mm,
    colback=white,before upper=\strut},
  title=#2,#1
}

\definecolor{custom-blue}{rgb}{0,0,0}

\newcommand{\tool}{\textsc{FuzzWise}\xspace}

\newboolean{showcomments}
\setboolean{showcomments}{true}
\ifthenelse{\boolean{showcomments}}
 { \newcommand{\mynote}[2]{
      \fbox{\bfseries\sffamily\scriptsize#1}
        {\small$\blacktriangleright$\textsf{\emph{#2}}$\blacktriangleleft$}}}
        { \newcommand{\mynote}[2]{}}

\usepackage{tikz}

\newcommand*\blackcircled[2][1.5em]{\tikz[baseline=(char.base)]{
            \node[shape=circle, fill=black, text=white, draw, minimum size=#1, inner sep=0] (char) {#2};}}

\newcolumntype{L}[1]{>{\raggedright\arraybackslash}p{#1}}

\newcommand{\code}[1]{{\footnotesize\texttt{#1}}}
\usepackage{amsthm}
\definecolor{dkgreen}{rgb}{0,0.6,0}
\definecolor{gray}{rgb}{0.5,0.5,0.5}
\definecolor{lightgray}{rgb}{211, 211, 211}
\definecolor{mauve}{rgb}{0.58,0,0.82}

\definecolor{custom-red}{rgb}{1,0,0}

\definecolor{c1}{HTML}{f4cccc}
\definecolor{c2}{HTML}{f5cdcd}
\definecolor{c3}{HTML}{fffcfc}
\definecolor{c4}{HTML}{ffffff}
\definecolor{c5}{HTML}{ffffff}
\definecolor{c6}{HTML}{fffdfd}
\definecolor{c7}{HTML}{f5cfcf}
\definecolor{c8}{HTML}{fffbfb}
\definecolor{c9}{HTML}{ffffff}
\definecolor{c10}{HTML}{fffdfd}
\definecolor{c11}{HTML}{fefafa}
\definecolor{c12}{HTML}{fef7f7}
\definecolor{c13}{HTML}{ffffff}
\definecolor{c14}{HTML}{fffefe}
\definecolor{c15}{HTML}{ffffff}
\definecolor{c16}{HTML}{fefafa}
\definecolor{c17}{HTML}{fdf3f3}
\definecolor{c18}{HTML}{fffefe}
\definecolor{c19}{HTML}{fdf5f5}
\definecolor{c20}{HTML}{ffffff}

\begin{document}


\title{{\tool}: Intelligent Initial Corpus Generation for Fuzzing}


\author{Hridya Dhulipala}
\orcid{0009-0001-4474-2984}
\affiliation{%
  \institution{University of Texas at Dallas}
  \city{Dallas}
  \country{USA}
}
\email{hridya.dhulipala@utdallas.edu}

\author{Xiaokai Rong}
\orcid{0009-0000-8457-8528}
\affiliation{%
  \institution{University of Texas at Dallas}
  \city{Dallas}
  \country{USA}
}
\email{xiaokai.rong@utdallas.edu}

\author{Aashish Yadavally}
\affiliation{
  \institution{University of Texas at Dallas}
  \city{Dallas}
  \country{USA}
}
\email{aashish.yadavally@utdallas.edu}

\author{Tien N. Nguyen}
\orcid{0009-0006-7962-6090}
\affiliation{%
  \institution{University of Texas at Dallas}
  \city{Dallas}
  \country{USA}
}
\email{Tien.N.Nguyen@utdallas.edu}

\setcopyright{none}

\settopmatter{printacmref=false, printfolios=false}

\begin{abstract}
In mutation-based greybox fuzzing, generating high-quality input seeds
for the initial corpus is essential for effective fuzzing. Rather than
conducting separate phases for generating a large corpus and
subsequently minimizing it, we propose {\tool} which integrates them
into one process to generate the optimal initial corpus of seeds
(ICS). {\tool} leverages a multi-agent framework based on Large
Language Models (LLMs). The first LLM agent generates test cases for
the target program. The second LLM agent, which functions as a
predictive code coverage module, assesses whether each generated test
case will enhance the overall coverage of the current corpus. The
streamlined process allows each newly generated test seed to be
immediately evaluated for its contribution to the overall
coverage. {\tool} employs a predictive approach using an LLM and
eliminates the need for actual execution, saving computational
resources and time, particularly in scenarios where the execution is
not desirable or even impossible. Our empirical evaluation
demonstrates that {\tool} generates significantly fewer test cases
than baseline methods. Despite the lower number of test cases, {\tool}
achieves high code coverage and triggers more runtime errors compared
to the baselines. Moreover, it is more time-efficient and
coverage-efficient in producing an initial corpus catching more errors.

  

\end{abstract}







\maketitle

\section{Introduction}

Mutation-based 
Fuzzing~\cite{fuzzguard,hongfuzz,kosta2016Conti} is a
dynamic approach to identify bugs
by exposing a target program to a high volume of random
inputs, with the goal of triggering runtime errors or crashes. The fuzzing process begins with an initial corpus of seed inputs, which are subjected to mutation and transformation. These mutated seeds generate new test cases that are executed against the target program to trigger anomalies, crashes, or failures. Effective fuzzing relies heavily on the quality of the initial corpus, as high-quality seeds significantly impact the fuzzer's performance ~\cite{klees2018evaluating,rebert2014opt,skyfire,moonlight}. If the initial corpus consists of seeds that explore similar paths or produce similar behaviors in the program, the fuzzing may become inefficient, as it may not effectively cover diverse areas.

To construct a high-quality initial corpus of seeds for the mutation
phase of a fuzzer, various approaches have been
employed~\cite{herrera2021seed,moonshine,afl}. Some generate test
cases randomly, while others use existing test suites, manually create
test inputs, real-world inputs, or example code snippets.
In several cycles of fuzzing, a full corpus is generated by
feeding the initial set of seed inputs into the fuzzer, which then
generates additional test cases through mutation,
expanding the corpus with new and varied inputs.
Since this full corpus is typically large and inefficient, a {\em corpus
minimization algorithm} (also called
{\em distillation})~\cite{moonlight,aflplus,herrera2021seed,moonshine,afl}
is applied to {\em automatically select a subset of seeds that
  is the smallest possible while still triggering the same
code coverage range as the full corpus}.
This smaller corpus is referred to as {\em the initial
  fuzzing corpus of seeds} or {\em the initial corpus of seeds}
(ICS). 


To achieve a higher bug yield through fuzzing, Hayes {\em et al.}~\cite{moonlight} identify properties that are desirable for the seeds forming the initial corpus.
The first property (\blackcircled[1em]{{\tiny P1}}) of the ICS is the {\em maximization
  of code coverage} of target behaviors: all the seeds in the initial
corpus should collectively span the wide range of observable behaviors
of the target program. All the fuzzers typically approximate this with
{\em code coverage}, {\em i.e.}, seeds should cover as much source code as
possible. Greybox fuzzers~\cite{afl,aflplus} compute code coverage by executing the target programs and employing lightweight code instrumentation to collect coverage data, which is later used during the minimization process.
Some fuzzers~\cite{afl} instrument edge transitions between basic
blocks of code to collect edge coverage. The second property
(\blackcircled[1em]{{\tiny P2}}) of the ICS is the {\em minimization
  of the total size of the corpus}. This helps reduce the mutation
search space in the fuzzing process. The third property
(\blackcircled[1em]{{\tiny P3}}) for an ICS is the {\em minimization
  of the redundancy/duplication} of seeds.  Fuzzing multiple seeds
with the same coverage is inefficient. The fourth property
(\blackcircled[1em]{{\tiny P4}}) is {\em the minimization of the sizes
  of the seeds}, as smaller seeds are preferable for reducing I/O
overhead on the storage system~\cite{moonlight}. However, this
property is not relevant in scenarios where input values themselves
are utilized as test cases.


In this work, we propose {\tool}, a novel predictive approach to {\em
  constructing a high-quality initial corpus of seeds} (ICS) for
mutation-based, coverage-guided fuzzers. First, instead of having two
dinstint phases—one for generating a large or full corpus and another
for minimizing it— we integrate them into one process of generating
the seeds in ICS to satisfy the above properties. {\tool} leverages a
Large Language Model (LLM)-based multi-agent framework in which the
first LLM agent generates test cases for the target program and the
second LLM agent, a predictive code coverage module, predicts whether
a generated test case would contribute to an increase in the total
coverage of the current corpus. If so, we add it into the ICS with the
goal of covering more un-tested areas. Otherwise, we prompt the
LLM-based test-case-generation agent for an alternate test input to
cover more un-tested areas in the subsequent cycle.  Such an
orchestration ensures an efficient exploration of mutation search
space to construct an ICS with higher coverage
(\blackcircled[1em]{{\tiny P1}}--\blackcircled[1em]{{\tiny P2}}),
while providing more control on the deduplication of seeds
(\blackcircled[1em]{{\tiny P3}}).  Finally, when the time limit is
reached or a total coverage reaches 100\%, the process stops and the
final ICS is obtained.

{\tool} represents a significant departure from traditional ICS construction methodologies. First, by integrating the generation and evaluation phases into a continuous process, {\tool} allows for ongoing improvement in code coverage throughout each cycle. This unified approach enables each newly generated test case to be immediately assessed for its impact on overall coverage. If a test case enhances coverage, it is incorporated into the corpus, thereby optimizing coverage incrementally. This dynamic strategy ensures that every test case is designed to maximize coverage from the beginning, unlike conventional methods that rely on a fixed initial corpus. 
Second, {\tool} eliminates the need for traditional instrumentation and execution of the target program. Instead of executing the target program to collect code coverage data, {\tool} utilizes a predictive model, CodePilot~\cite{forge24}, which predicts coverage without actual execution. This approach conserves computational resources and time, particularly in environments requiring complex setups or in situations that actual execution is not preferrable or even impossible, and facilitates the seamless integration of test case generation and coverage estimation.

We conducted several experiments to evaluate {\tool}. Our result on
the FixEval dataset~\cite{haque2023fixeval} shows that within a fixed
time frame, {\tool} produces higher-quality ICS compared to current
state-of-the-art corpus generation and minimization techniques.  It
achieves this by generating significantly fewer test cases (42.1\%
fewer) while delivering greater code coverage (10.71\% higher). The
ICSes produced by {\tool} are notably more effective at detecting
runtime errors than that of baseline methods.
Additionally, the ICS from {\tool} serves as a superior starting point
for mutation-based fuzzing, resulting in a {\color{custom-blue}{100\%}}
relative increase in the detection of runtime errors compared than the
ICS produced by the baselines. We also showed that {\tool} is more
time efficient than the baselines in producing an ICS with the highest
code coverage, {\em i.e.}, its prediction time is less than the
execution time and the minimization time of the baselines to obtain
the corpora with the same coverage. {\tool} is also more efficient in
terms of corpus' size and code coverage in detecting more runtime
errors/exceptions.
In brief, the key contributions of this paper include:


{\bf 1. {\tool}:} a novel predictive algorithm for the construction of
a high-quality initial corpus of test seeds for mutation-based fuzzing.


{\bf 2. A Tandem of complementary LLM-based Agents}: the LLM
test generation improves better recall, while the LLM coverage
prediction helps better precision in building a high-quality ICS.

{\bf 3. Extensive Empirical Evaluation:} Our results demonstrate that {\tool} is
more effective and efficient than the baseline models. Data
and code is available at~\cite{fuzzwise-website}.

\section{Motivation}
\label{sec:motiv}

\subsection{Motivating Example}



In this section, we utilize an example to illustrate the problem and
motivate our solution. Fig.~\ref{fig:motiv} displays a Java program
from our experimental dataset.

\begin{wrapfigure}{r}{2.6in}
\small
	\centering
	\lstset{
		numbers=left,
		numberstyle= \tiny,
		keywordstyle= \color{blue!70},
		commentstyle= \color{red!50!green!50!blue!50},
		frame=shadowbox,
		rulesepcolor= \color{red!20!green!20!blue!20} ,
		xleftmargin=1.5em,xrightmargin=0em, aboveskip=1em,
		framexleftmargin=1.5em,
                numbersep= 5pt,
		language=Java,
    basicstyle=\scriptsize\ttfamily,
    numberstyle=\scriptsize\ttfamily,
    emphstyle=\bfseries,
                moredelim=**[is][\color{red}]{@}{@},
		escapeinside= {(*@}{@*)}
	}
\begin{lstlisting}[]
import java.util.Arrays;
import java.util.Scanner;
public class Main {
  public static void main(String[] args) {
    Scanner sc = new Scanner(System.in);
       int n = sc.nextInt();
       int[] a = new int[n];
       for (int i = 0; i < n; i++) {
         String type = sc.next();
           if (type.equals("S")) {
             ...
           } else ...
         }
       Arrays.sort(a);
       for (int i = 1, j = 0; i <= 52; i++) {
         if (i <= 13) {
            if (a[j] == 100 + i) {
              j++;
            } else {
              ...
            }...
}            
\end{lstlisting}
\vspace{-16pt}
\caption{An Example of a Target Java Program}
\label{fig:motiv}
\end{wrapfigure}

We used a coverage-guided fuzzing tool, JQF-AFL~\cite{JQF} on
Fig.~\ref{fig:motiv} to generate test cases.  First, JQF randomly
generated a set of 5 test cases (inputs) as seeds. It then mutated
these seed test cases to produce a full corpus of 65 test cases, which
account for a total of 97.36\% statement coverage. To reduce this full
corpus, JQF invoked its corpus minimization (\code{JQF-AFL-cmin}),
which reduced the corpus to an initial set of 46 test cases (ICS)
while maintaining the same coverage.  We then ran the test cases in
the ICS from JQF, in which none were able to detect any runtime
error. To continue, JQF randomly selected a new set of test
seeds. Note that during the minimization process, JQF performs
lightweight executions to measure the code coverage for the test cases
in the full corpus. Thus, a notable proportion of the generated test
cases possess code coverages not contributing to the increasing total
coverage, leading to inefficiencies and the squandering of runtime
resources in selecting the seeds for ICS.

Importantly, even in the full corpus or in the minimized ICS, despite
of many of the generated test cases,
none was successful in unveiling 
\code{ArrayIndexOutOfBoundException} present in the code, as listed in
the ground truth in our dataset. This poses a considerable risk of
overlooking or inadequately addressing certain runtime errors that
may manifest during program execution with incorrect variables'
values. Moreover, some test cases were
redundantly generated, further exacerbating inefficiencies and failing
to contribute to the enhancement of the quality of ICS in later
iterations.

\subsection{Key Ideas}\label{sec:key-ideas}

We design {\tool} with the following key ideas:

\subsubsection{A Unified Process of Constructing the Initial Corpus of Seeds (ICS)}
State-of-the-art approaches for constructing the initial corpus of
seeds (ICS) typically involve generating a full corpus of test cases,
following which, corpus distillation or minimization techniques are applied to
select the most valuable test cases for the ICS (with highest possible total
coverage). 
%
In doing so, the minimization process does not actively seek to increase or improve the code coverage of the ICS, as the goal is purely to remove redundant test cases from the generated full corpus.
Moreover, once the full corpus is established, the construction of the minimized corpus does not contribute to additional code coverage. It does not guide or improve the selection of subsequent test cases during the minimization process, leading to inefficiencies.
In the example, the minimization
would not contribute to a higher coverage than \textcolor{black}{97.36\%}, which is the
coverage for the full corpus.


Our first key idea in {\tool} is a unified process for ICS
construction. This process {\em integrates test case generation and
  upfront quality assessment through code coverage prediction}. The
first component of {\tool} utilizes a Large Language Model (LLM)-based
agent to generate test cases for the target code. LLMs have
demonstrated their ability to significantly expand the exploration of
the test case space, which enhances the effectiveness of
fuzzing~\cite{fuzz4all}. However, this broad exploration can sometimes
compromise the precision of the generated test cases. To address this,
we leverage our second component, an LLM-based agent named
CodePilot~\cite{forge24}, to assess the quality of the generated test
cases by predicting their contribution to the corpus' code coverage.


Our approach includes a feedback loop between the two LLM agents. If
CodePilot predicts that a generated test case will not increase the
total code coverage of the corpus, it feeds this information back to
the test case generation agent. This provides a signal for the generation process to
adapt and improve continuously, ensuring that each new test case has a
higher likelihood of enhancing code coverage. If the test case is
deemed valuable, it is added to the current corpus, and the unified
ICS construction process continues, thus systematically covering
previously untested areas of the code with the new test cases.
In the above example, {\tool} successfully generated 3 test cases in
the ICS with 100\% statement coverage, which correctly detected the
runtime error.
This shows an improvement compared to the generation of 46 test
cases by JQF with none runtime error detected.



\subsubsection{Predictive Code Coverage as a Guidance for Test Case Generation} 
State-of-the-art approaches rely on code coverage as an indicator of
test case quality, using it to assess how well test cases cover
program behaviors. Code coverage serves as the primary guide for the
minimization process, where the goal is to reduce the full corpus to
an ICS by eliminating test cases that trigger similar behaviors, as
indicated by similar code coverage. This reduction minimizes wasted
effort in fuzzing, including the overhead of environment setup and
test execution, even though existing fuzzers employ lightweight code
instrumentation to gather this coverage information.

In contrast, {\tool} employs CodePilot~\cite{forge24} for code
coverage prediction, which, while still assessing test case quality,
shifts the guidance from corpus minimization to test case
generation. By predicting code coverage without requiring actual
execution, it optimizes the ICS construction process, making it more
efficient and reducing the resource demands typically associated with
environment setups and test executions. This predictive approach
allows {\tool} to dynamically guide test case generation, ensuring that
each test case is more likely to contribute to meaningful coverage
improvements, all while maintaining a streamlined and
resource-efficient process.


\subsubsection{A Tandem of LLM-based Test Generation and LLM-based Code Coverage Prediction} In {\tool}, we utilize a tandem of LLMs, each serving distinct roles: one for test generation from the source code and the other for code coverage prediction.
The main objective of the first LLM is to expand the exploration of
the test case space for the given source code, thereby expecting to
increase recall. However, this may come at the expense of precision in
the generated test cases. To counterbalance this trade-off, the second
LLM used with CodePilot~\cite{forge24} prompting for code coverage
prediction aims to enhance precision by assessing the quality of the
generated test cases. Morever, any duplicate test cases generated by
the first LLM are discarded, to avoid redundancy.
\section{Predictive Coverage-Guided Fuzzing Overview}
\label{sec:framework}

\begin{figure}[t]
\begin{center}
\includegraphics[width=5.4in]{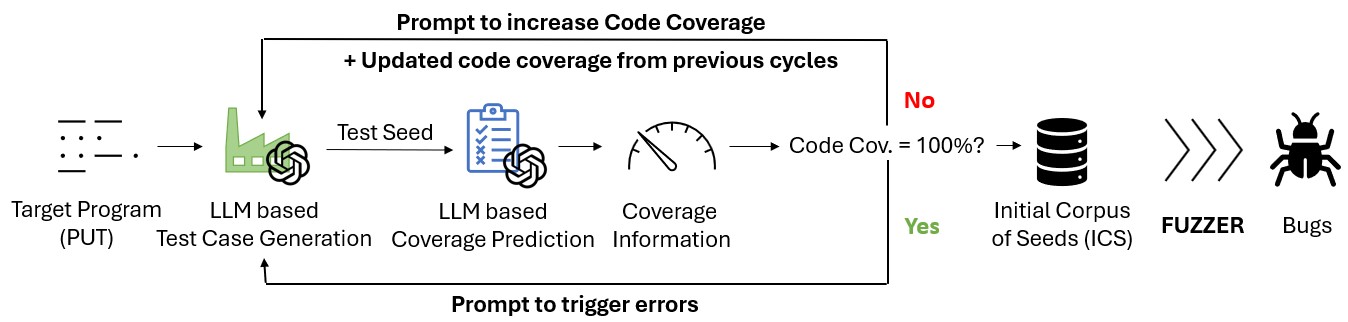}
\vspace{-9pt}
\caption{{\tool}: Predictive Coverage-Guided Construction of Initial Corpus of Seeds}
\label{fig:fuzzwise}
\end{center}
\end{figure}

Fig.~\ref{fig:fuzzwise} shows {\tool}'s overall workflow. The core of
{\tool} is a tandem of two Large Language Models (LLMs): 1) the
LLM-based Test Case Generator (TCG), and 2) CodePilot~\cite{forge24},
the LLM-based code coverage predictor. Both work collaboratively in an
iterative fashion to build the ICS.

First, the LLM-based TCG module examines the source code of the
program under test and generates a test case accordingly. This test
case is fed into CodePilot~\cite{forge24} for the estimation of its
quality in terms of code coverage. The new test case will be collected
if it is estimated to explore the new area of code (i.e., the total
code coverage of the corpus with the new test case is estimated to be
higher than the coverage of the current one). This test generation
cycle continues until a total code coverage of 100\% is achieved. In
the next cycle, we also include the updated code coverage to the
prompt to the LLM-based TCG. If the code coverage reaches 100\%, the
LLM-based TCG switches to generating test cases specifically designed
to trigger errors in the program under test (PUT). If the time limit
is reached, and the coverage does not reach 100\%, {\tool} also
switches to the LLM-based TCG. Test cases that are duplicates or have
already been added to the corpus are discarded, and the LLM is
prompted to generate a new, distinct test case. Test cases are added
to the ICS only if they either cover
previously unexplored code or trigger errors. The process
continues until the specified time limit is reached and the resulting ICS
is returned for further fuzzing.




\section{{\tool} Algorithm}
\definecolor{gray}{rgb}{0.5,0.5,0.5}
\definecolor{mauve}{RGB}{127,0,145}
\definecolor{lightgray}{gray}{0.97}


\begin{wrapfigure}{r}{2.8in}
\small
\centering
\lstset{
  language=Java, caption= FuzzWise Workflow Algorithm, float, label={algo}, morekeywords={do, while, foreach,function,equals,and,or,in}, mathescape=true, escapechar=|,backgroundcolor=\color{lightgray},
	numbersep=-4pt,
	numberstyle=\tiny\color{gray},
	keywordstyle=\color{mauve},
        basicstyle=\scriptsize\sffamily,
        moredelim=**[is][\color{red}]{@}{@},
	escapeinside= {(*@}{@*)}
}
\begin{lstlisting}[]
    $Corpus$ = $\emptyset$ 
    $Cumulative\_cov$ = $\emptyset$ 
    $Errors$ = $\emptyset$ 
    $Stop$ = False 
    $Pred\_cov$ = $\emptyset$ 
    while (not $Stop$): 
        if ($Cumulative\_cov < 100\%$): 
            $T\_seed$ = (*@{\color{blue}{TCG\_LLM}}@*)($prompt_{IC}$) 
        else: 
            $T\_seed$ = (*@{\color{blue}{TCG\_LLM}}@*)($prompt_{TE}$) 
        $Pred\_cov$ = (*@{\color{orange}{COV\_LLM}}@*)($T\_seed$, $prompt_{CCP}$) 
        if ($Pred\_cov \cup Cumulative\_cov$ > $Cumulative\_cov$): 
            $Corpus$.extend($T\_seed$) 
            $Cumulative\_cov$ = update($Cumulative\_cov$,$Pred\_cov$)
        if (time-limit-reached): 
            $Stop$ = True 
    RemoveDuplicates($Corpus$) 
    return $Corpus$
\end{lstlisting}
\end{wrapfigure}

Listing~\ref{algo} shows the pseudo-code illustrating {\tool}
workflow. It begins by initializing the key variables (lines 1--5):
$Corpus$ to store generated test seeds in the initial corpus of seeds,
$Cumulative\_cov$ to track total coverage of the current corpus
(starting with an empty one), $Stop$ as a termination flag, and
$Pred\_cov$ for predicting coverage of the generated seed. The
iterative process continues until a time limit is reached (lines 6--16).

The decision-making process within each iteration is dynamic, with
branching based on the evaluation of predicted coverage and the
cumulative coverage. {\color{custom-blue} The Test Case Generation LLM, {\em
    $TCG\_LLM$}, is invoked (line 8) with a prompt to generate a test
  case aimed at either increasing code coverage or triggering errors,
  depending on whether the cumulative coverage has reached 100\%.  If
  the coverage is below 100\%, the TCG\_LLM continues to focus on
  generating test cases for uncovered code (line 8); otherwise, the
  TCG\_LLM shifts to generating error-detecting test cases (line
  10). Next, the predicted code coverage of the generated test case is
  estimated by the Code Coverage Prediction LLM, {\em CodePilot~\cite{forge24}}
  (line 11). If the predicted coverage of the new test case contributes to
  a higher cumulative coverage, it will be added to the $Corpus$, and
  the cumulative coverage is adjusted accordingly (lines 12–14). If
  the new test case is estimated not contributing to additional coverage or
  error detection, it is discarded. The process continues until the
  time limit is reached (line 15), ensuring FuzzWise operates
  efficiently within a given time. Upon conclusion, duplicate
  test cases are removed from the corpus (line 17).}

In summary, {\tool} employs dynamic test generation strategies, code coverage prediction, and cumulative coverage tracking to effectively and progressively build the ICS.

\section{LLM-based Test Generation and LLM-based Predictive Code Coverage}
\label{sec:llm}

While the previous two sections present the overview and workflow in
{\tool}, this section presents the design of the interactions via
prompting between the LLMs used for test case generation and coverage
prediction. These prompting steps correspond to the API calls to the
LLM at lines 8 and 10 of Listing~\ref{algo}, and the call to coverage
prediction LLM, CodePilot~\cite{forge24} at line 11 of
Listing~\ref{algo}.

\subsection{The Test Case Generation Module}

The Test Case Generation module is operated by prompting
GPT-4~\cite{chatGPT} to generate test cases in a specific format.
We implement two distinct  distinct prompts, each serving a different purpose. One prompt is designed to generate test
cases targeting various vulnerable areas (i.e., increasing the code
coverage), while the other prompts focuses on each area individually
to reveal errors/exceptions.

\subsubsection{Prompt for Test Case Generation with the Goal of Increasing Code Coverage}

As {\tool} functions in a coverage-guided fuzzing framework,
increasing coverage stands as one of primary objectives. This section of our
prompt is directed to create test cases likely to {\em increase
  coverage by targeting the portions of code that are uncovered}
(the prompt is depicted in Fig.~\ref{fig:prompt-coverage-increasing}). In the
prompt, the LLM receives {\em the code covered by all the seeds in the
  current corpus, denoted by `$>$' and `!'  for covered and uncovered
  lines}, respectively. {\tool} instructs the LLM (line 8 of
Listing~\ref{algo}) to generate test cases while accounting for a
variety of input scenarios, including negative, positive, zero, and
maximum values. This ensures that a diverse range of inputs is
considered.





\subsubsection{Prompt for Test Case Generation with the Goal of Detecting Runtime Errors}

The second strategy (line 10 of Listing~\ref{algo}) is
designed to prompt the LLM towards generating test cases aimed at
catching runtime errors and
exceptions. Fig.~\ref{fig:prompt-error-triggering} illustrates the
structures and instructions included in our prompt for this strategy.
The objective is to
generate test cases to explore diverse facets of runtime exceptions
comprehensively. We expect that this strategy complements the
coverage-increasing test case generating strategy to efficiently
produce high-quality ICS for fuzzing.



\subsection{LLM-based Code Coverage Prediction}
\label{sec:codepilot}

\begin{figure}[t]
  \centering
	\lstset{
		numbers=left,
		numberstyle= \tiny,
		keywordstyle= \color{blue!70},
		commentstyle= \color{red!50!green!50!blue!50},
		frame=shadowbox,
		rulesepcolor= \color{red!20!green!20!blue!20} ,
		xleftmargin=1.5em,xrightmargin=0em, aboveskip=1em,
		framexleftmargin=1.5em,
                numbersep= 5pt,
		language=Java,
    basicstyle=\scriptsize\ttfamily,
    numberstyle=\scriptsize\ttfamily,
    emphstyle=\bfseries,
                moredelim=**[is][\color{red}]{@}{@},
		escapeinside= {(*@}{@*)}
	}
\begin{lstlisting}[caption = ]
(*@{\color{orange}{PROMPT} on Test Case Generation for Increasing Code Coverage@*)
Generate a single test case for a Java program to cover uncovered lines of code denoted by '!'. Provide only the test input without explanations. Consider various conditions, edge cases, and typical use cases.
Ensure the test case input is in the following format: ...
Test Case Input:
<input 1> ...
Code Coverage of Current Corpus: ...
\end{lstlisting}
\vspace{-20pt}
\caption{Prompt for Test Case Generation with the Goal of Increasing Code Coverage}
\label{fig:prompt-coverage-increasing}
\end{figure}

In our study, we utilize CodePilot~\cite{forge24}, an LLM-based
approach for predicting code coverage of a program. CodePilot uses
step-by-step reasoning to guide the LLM in analyzing path constraints
for various execution paths. It helps predictively
execute the code based on specific inputs and determine code
coverage. Unlike planning in robotics or
NLP~\cite{zhuang2024toolchain,hu2023avis,yao2023react,prasad2023adapt},
which relies on the proficiency of the LLM in text comprehension,
CodePilot's step by step reasoning for code coverage is rooted in
program semantics and analysis, incorporating the complexity of code
execution. Consequently, CodePilot requires a guided exemplary plan
that discerns various types of statements and treats them differently
to enhance its understanding of code execution. More details can be
found in~\cite{forge24}.
The interaction between {\tool} and CodePilot is expressed at line 11
of Listing~\ref{algo}. CodePilot is called to estimate the code
coverage for each test seed, and provides an output that indicates
which lines of code are (un)covered. {\tool}
then parses this output to calculate the predicted coverage.

\begin{figure}[t]
  \centering
	\lstset{
		numbers=left,
		numberstyle= \tiny,
		keywordstyle= \color{blue!70},
		commentstyle= \color{red!50!green!50!blue!50},
		frame=shadowbox,
		rulesepcolor= \color{red!20!green!20!blue!20} ,
		xleftmargin=1.5em,xrightmargin=0em, aboveskip=1em,
		framexleftmargin=1.5em,
                numbersep= 5pt,
		language=Java,
    basicstyle=\scriptsize\ttfamily,
    numberstyle=\scriptsize\ttfamily,
    emphstyle=\bfseries,
                moredelim=**[is][\color{red}]{@}{@},
		escapeinside= {(*@}{@*)}
	}
\begin{lstlisting}[caption = ]
(*@{\color{orange}{PROMPT} on Test Case Generation for Detecting Errors and Exceptions@*)
Generate a single test case without any explanation to raise the following scenarios in a Java program:
InputMismatchException: Provide an input value that whose data type is different than the one specified. 
ArithmeticException: Test cases that could raise arithmetic exceptions include division by zero, overflow, underflow, or attempt to perform invalid operations, e.g., taking the square root of a negative number.
NullPointerException: Create a scenario where a variable is explicitly set to null before usage.
NumberFormatException: Input a value that cannot be parsed to the expected data type (a non-numeric string).
ArrayIndexOutOfBoundsException or IndexOutOfBoundsException: Design input values that lead to accessing array or list indices beyond their bounds.
(Other types of runtime errors and exceptions)
Ensure the test case input is in the following format:
Test Case Input:
<input 1>
<input 2>...
Generate a single test case without providing an explanation for the below Java program: ...
\end{lstlisting}
\vspace{-20pt}
\caption{Prompt for Test Case Generation with the Goal of Detecting Runtime Errors and Exceptions}
\label{fig:prompt-error-triggering}
\end{figure}

\section{Empirical Evaluation}
\label{sec:eval}

For evaluation, we seek to answer the following questions:

\noindent {\bf RQ1. [Effectiveness of {\tool} in Construction of
    Initial Corpus of Seeds].} How effective is {\tool} in
constructing the initial corpus of seeds (ICS) for mutation-based
fuzzing?

\noindent {\bf RQ2. [Efficiency of {\tool} in Construction of Initial
    Corpus of Seeds].} How efficient is {\tool} in constructing the initial corpus of seeds (ICS) for mutation-based fuzzing?


\noindent {\bf RQ3. [Evaluation on Prompting Strategy in Targeting to Detect Errors]} How effective is {\tool}'s prompting strategy in the \emph{TCG LLM} for detecting errors?

\noindent {\bf RQ4. [Usefulness in Fuzzing].} How useful is {\tool} in working with a fuzzer?







\subsubsection*{Dataset}

In this study, we employed the FixEval
dataset~\cite{haque2023fixeval}, a benchmark comprising real-world
erroneous code submissions to competitive programming challenges in
the AtCoder~\cite{atcoder} and Aizu Online Judge~\cite{aizu}
platforms. We randomly selected 50 examples, each of Java and Python submissions
from this dataset, focusing on variations in code size and branching
factor.
Furthermore, following the guidelines outlined in~\cite{nasa} for path
testing based on structural complexity, we selected instances with
cyclomatic complexities ranging from 10 to 20.  This selection ensures
a representative sample of real-world code while addressing the token
constraints of our underlying LLM, GPT-4~\cite{chatGPT}.



\section{Effectiveness in Generating Initial Corpus of Seeds (RQ1)}
\label{sec:rq1}


\subsection{Experimental Methodology}

\subsubsection{Baselines}


To evaluate {\tool}'s effectiveness in generating the initial corpus of seeds, we established several baselines in Java and Python. For the Java-specific baselines, we selected the widely used fuzzing framework {\em JQF-AFL}~\cite{JQF}, along with its corresponding corpus minimization tool, {\em JQF-AFL-cmin}. Similarly, for the Python-specific baselines, we opted for {\em Python-AFL}~\cite{py-afl} and its corresponding minimization algorithm, {\em Py-AFL-cmin}.

{\em Baseline 1}: Here, we used the fuzzing frameworks ({\em i.e.}, {\em JQF-AFL} and {\em Python-AFL}) to randomly generate the initial corpus of seeds (ICS) directly. Both create an initial set of \underline{five} random test seeds, which serve as the starting point for the fuzzing process. These are generated with the aim of covering a broad spectrum of potential inputs, helping the fuzzer explore diverse execution paths.


{\em Baseline 2}: Next, to represent a typical procedure to produce an ICS by applying minimization after randomly generating a full corpus, we used the fuzzing framework to generate \underline{50} seeds and then applied the corresponding distillation tools to minimize it and produce the ICS.

{\em Baseline 3}: To illustrate a common method for creating an ICS through random seed generation, mutation, and corpus minimization, we used both fuzzers to randomly generate a small corpus of \underline{5} test seeds. Then, we fuzzed these test seeds on the Program Under Test (PUT) until a
stopping criterion is met, such as a time limit or a specified number
of seeds in the corpus. Following this, the corresponding corpus minimization algorithm was applied to filter and minimize the corpus.

\subsubsection{Procedure}
To ensure a consistent evaluation, we maintained an environment where
each experimental setup was given a fixed runtime of 5 minutes on each code snippet
in the dataset. This ensures that all models were assessed under the
same time frame. Before running the tests, we ensured that the
environment was reset for each execution to prevent any residual data
or influence from previous runs. Except for Baseline 1, which used 5 random test seeds, each model generated an initial corpus of seeds (ICS) within the time limit.
For Baseline 2, this duration includes the random generation of 50
test seeds, their execution by the fuzzing framework to record code coverage, and the subsequent minimization to produce the initial
corpus. For Baseline 3, the 5-minute period includes both the random
generation of the 5 initial seeds and fuzzing, after which the
full corpus was minimized using the corresponding corpus minimization algorithm to derive the ICS. For
{\tool}, the 5-minute time limit involved the generation of each test
seed by the TCG LLM, followed by the prediction of the coverage on the
PUT using CodePilot.



\subsubsection{Metrics}
In this study,
we measure the number of unique bugs/faults triggered by the ICS.





\subsection{Empirical Results}
\label{sec:rq1-results}

\begin{figure}[t]
\begin{center}
\includegraphics[width=3.3in]{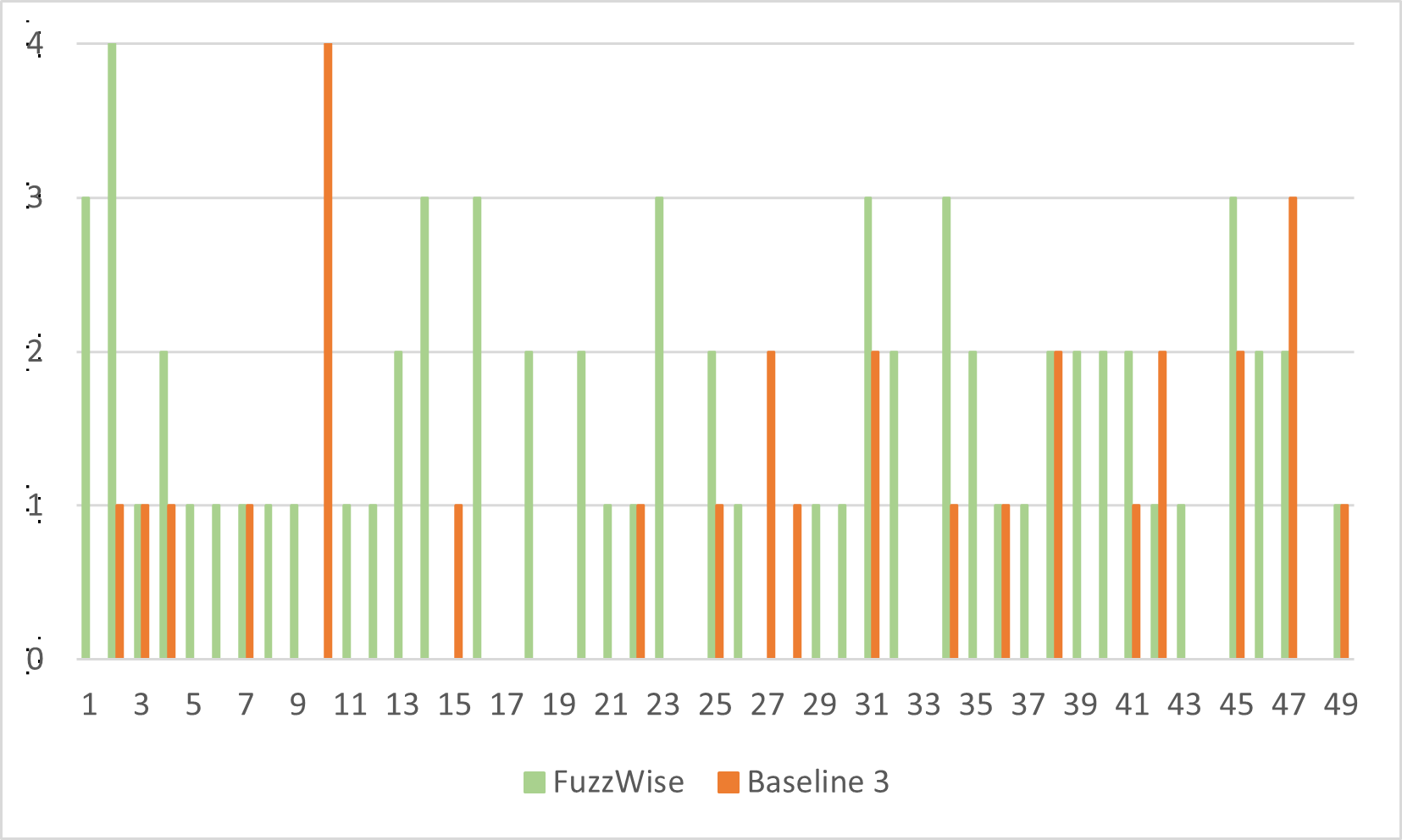}
\vspace{-9pt}
\caption{Number of Unique Errors Triggered for All \underline{Java} Programs (RQ1)}
\label{fig:rq1_errors}
\end{center}
\end{figure}



\subsubsection{Results for Java}\label{sec:rq1-results-java}
As shown in Fig.~\ref{fig:rq1_errors}, it is clear that {\tool} triggers more errors in the Java programs within the 5-minute time limit compared to all the Baselines 1--3.

To be specific, {\tool} outperforms Baseline 3, {\em i.e.}, where {\em JQF-AFL} fuzzes 5 randomly generated input seeds for 5 minutes before applying {\em AFL-cmin} for minimization. {\tool} (depicted in green) triggers an equal or greater number of bugs than Baseline 3 (represented in orange) in {\bf 38 out of 50} ({\em i.e. 76\%}) programs. This improved performance can be attributed to {\tool}'s targeted test-case generation, which takes advantage of the feedback provided by the code coverage prediction module (see Key Idea 1, Section~\ref{sec:key-ideas}) to explore better test case space, triggering more errors than the random generation, blind mutation and inefficient minimization strategy applied by JQF-AFL-cmin.



In addition, {\tool} detects fewer errors compared to
Baseline 3 in only {\em 6 out of 50} programs ({\em i.e.}, 12\%). 
This can be attributed to Baseline 3 generating a larger
pool of seeds during the fuzzing phase. After applying minimization,
this results in an average corpus size that exceeds that of
{\tool}. While these seeds are randomly generated, the larger corpus
size theoretically increases the chances of uncovering more
bugs. However, this happens infrequently due to the random and
unguided nature of the seed generation in Baseline 3.

Upon further inspection, we also observed that these 6 programs ({\em
  i.e.}, 17, 19, 24, 33, 44, and 48 in Fig.~\ref{fig:rq1_errors}) are,
on an average, more complex than the rest (cyclomatic complexity of
18, while that of the remaining 44 programs is 14). Despite this,
{\tool} exhibits an increasing coverage for these programs, indicating
that it possibly requires more time than the allocated time limit to
generate more bug-triggering test cases. It is important to highlight
that the lower performance on these 6 programs does not reflect the
overall performance on more complex programs, as these represent only
a subset ({\em 6 out of 15}) within this complexity range. In 6 out of
50 programs, neither {\tool} nor Baseline 3 manage to trigger any
bugs due to their high complexity.

Notably, both Baseline 1 (using 5 random test cases as ICS)
and Baseline 2 (starting with 50 test cases, then minimized to obtain
ICS) failed to detect any errors across the dataset. This outcome stems
from JQF's random seed generation mechanism not being effective enough
to produce error-triggering inputs in Baseline 1. Even in Baseline 2,
where some error-inducing seeds may exist, JQF's AFL-cmin algorithm,
used for corpus minimization, excludes these seeds from the resulting
corpus. Consequently, Baselines 1 and 2 are not shown in
Fig.~\ref{fig:rq1_errors} as they do not trigger any errors.



\subsubsection{Results for Python}

\begin{figure}[t]
\begin{center}
\includegraphics[width=3.4in]{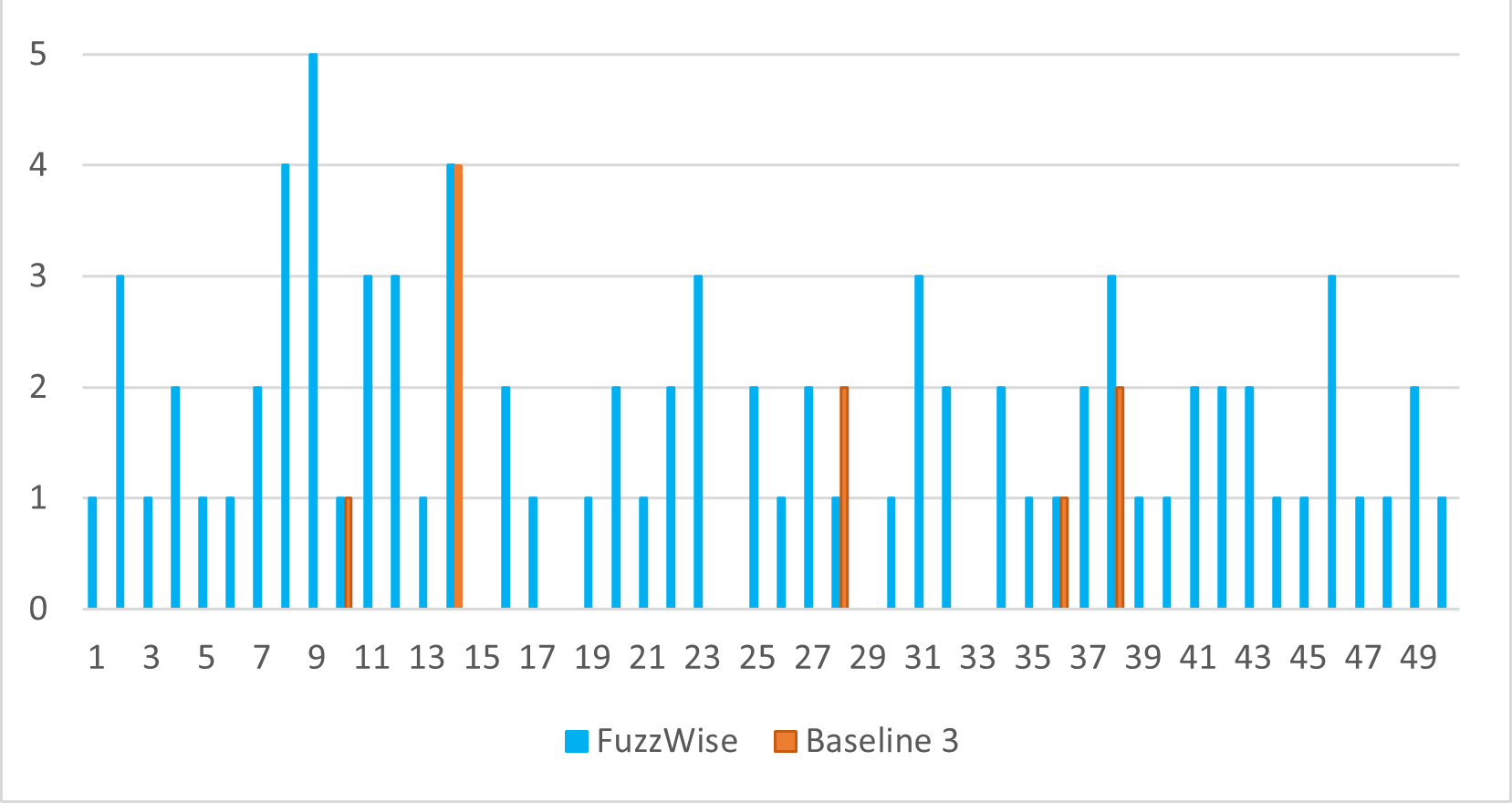}
\vspace{-9pt}
\caption{Number of Unique Errors Triggered for All \underline{Python} Programs (RQ1)}
\label{fig:rq1_python_errors}
\end{center}
\end{figure}

As shown in Fig.~\ref{fig:rq1_python_errors}, {\tool} significantly
outperforms all baselines in detecting errors in Python programs
during the 5-minute window. Consistent with the observations on Java programs (Section~\ref{sec:rq1-results-java}), 
both Baselines 1 and 2 failed to trigger
any errors across the dataset. This can be attributed to the
inefficiency of randomly generated seeds in Baseline 1 and the
limitations of {\em Py-AFL-cmin} in Baseline 2.
Compared to Baseline 3, {\tool} demonstrates a significantly better
performance. As depicted in Fig.~\ref{fig:rq1_python_errors}, {\tool}
(represented in blue) detects an equal or greater number of bugs than
Baseline~3 (represented in orange) in 44 out of 50 ({\em i.e.}, 88\%)
programs.  This improvement in performance is due to our prompting
strategy guiding the LLM to generate seeds to explore error-triggering
execution paths (see Section~\ref{sec:rq3} for more analysis).  In
contrast, Py-AFL relies on random input generation that may not
effectively target buggy areas.

Moreover, in {\em 5 out of the 50} {\em i.e.} ({\em 10\%}) programs,
neither {\tool} nor Baseline 3 manages to detect any bugs (e.g.,
programs 15, 18, 24, 29, and 33 in Fig.~\ref{fig:rq1_python_errors}).  The
failure in these cases is due to the significantly larger size and
complexity of these particular code snippets. Lastly, Baseline 3 finds one error more than {\tool} in a single program (program 28) out of the entire dataset. This rare observation is due to the larger corpus generated by Baseline 3.

\section{Efficiency in Generating the Initial Corpus of Seeds (RQ2)}
\label{sec:rq2}

\subsection{Experimental Methodology}





\subsubsection{Metrics} 
We use the same procedure and setup as in RQ1 (Section~\ref{sec:rq1}) to analyze {\tool}'s efficiency in generating ICSes.
To this end, we evaluate efficiency across three dimensions:

\begin{itemize}[topsep=0pt, leftmargin=10pt]
    \item {\em Errors per Seed (EPS)} is measured as the ratio of the average number of errors triggered by a generated ICS for a given PUT to the total number of seeds in the ICS. This metric assesses the model’s efficiency in generating error-triggering seeds, {\em i.e.}, a higher EPS indicates that a model can trigger more errors with fewer number of seeds.
    \item {\em Errors per Coverage (EPC)} is the ratio of the average number of errors triggered by the model within a fixed time to the cumulative code coverage achieved on the PUT by the generated ICS. EPC serves as an indicator of a model's efficiency in bug detection with respect to code coverage. Higher EPC values suggest that the generated ICS by a model can trigger more bugs with a smaller code coverage, making it a measure of balancing between coverage and error detection.
    \item {\em Errors per Time (EPT)}, time-efficiency, is computed as the average number of detected errors in a unit of time. A higher EPT value signifies that the generated ICS by a model is more efficient at triggering errors within a shorter time, demonstrating its promptness in bug detection.
\end{itemize}





\subsection{Empirical Results}
\subsubsection{Results for Java}

\begin{figure}[t]
\begin{center}
\includegraphics[width=3.4in]{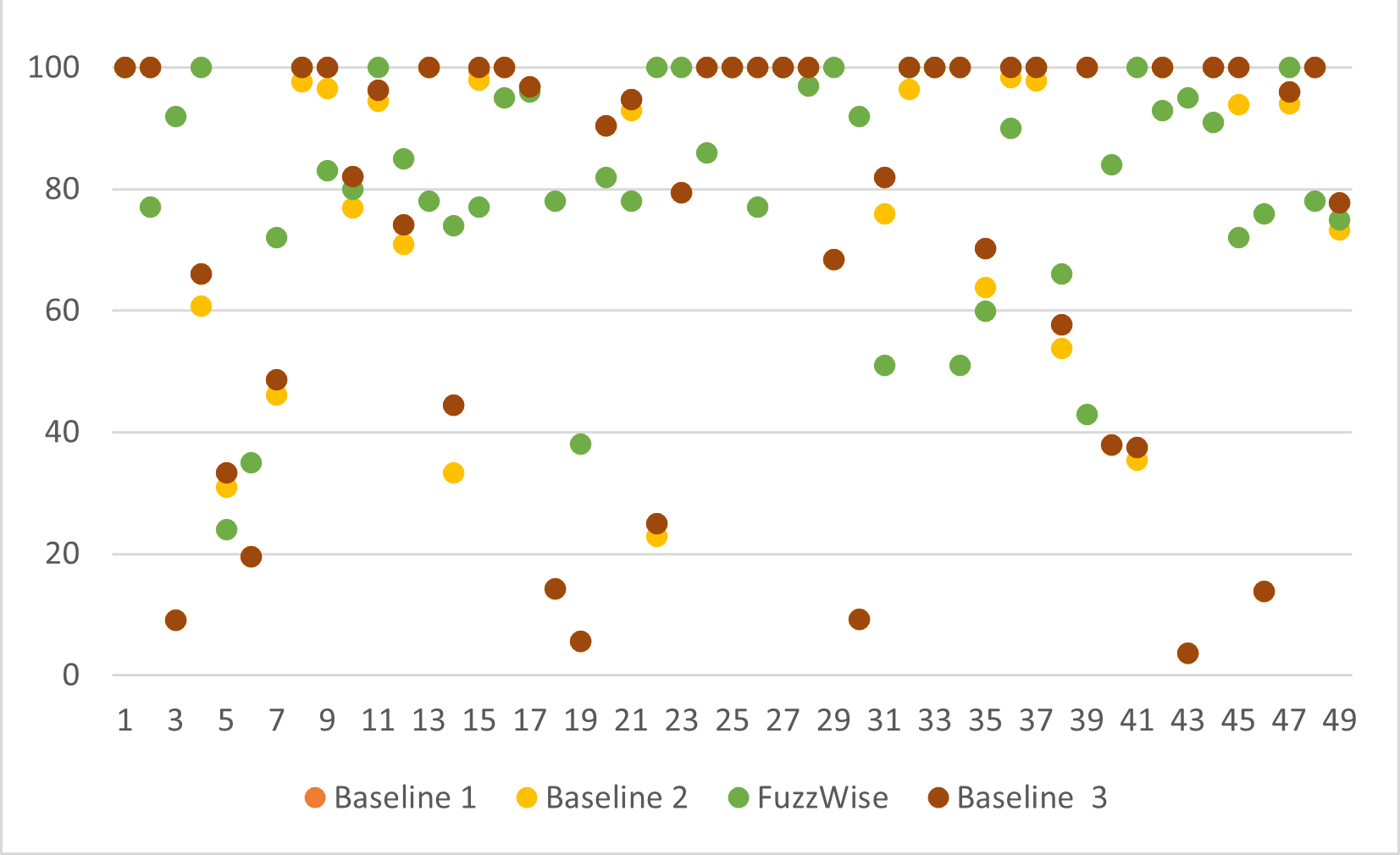}
\vspace{-9pt}
\caption{Coverages in All \underline{Java} Programs (RQ2)}
\label{fig:rq1_coverage}
\end{center}
\end{figure}

\begin{figure}[t]
\begin{center}
\includegraphics[width=3.4in]{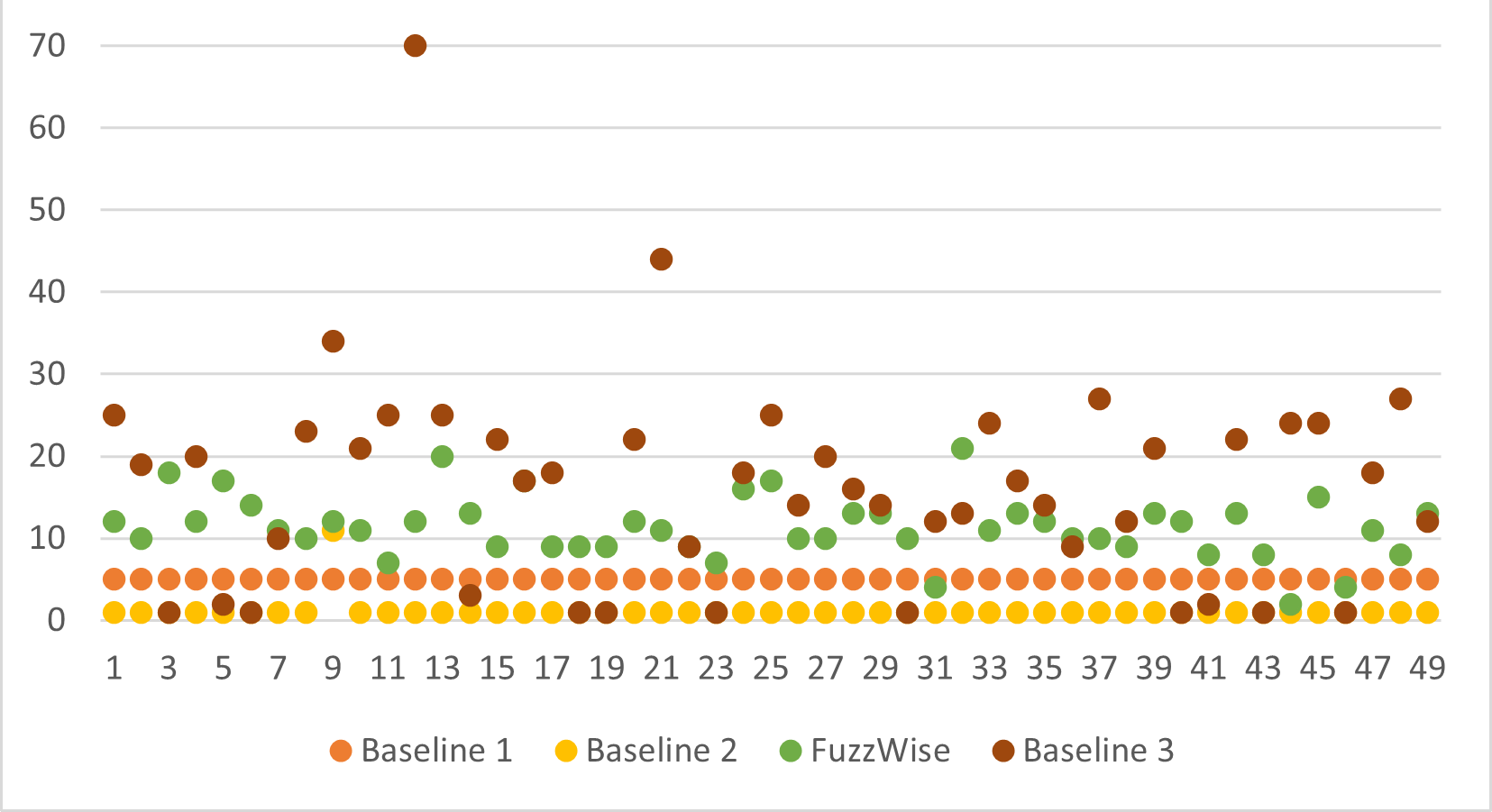}
\vspace{-9pt}
\caption{Corpus Sizes for All \underline{Java} Programs (RQ2)}
\label{fig:rq1_corpus_size}
\end{center}
\end{figure}

\begin{table}[t]
\small
\caption{Efficiency in Generating the Initial Corpus of Seeds for \underline{Java} (RQ2)}
\vspace{-6pt}
\label{tab:rq2-results}
\centering
\begin{tabular}{|l|c|c|c|}
\hline
& \multicolumn{1}{c|}{\begin{tabular}[c]{@{}c@{}}Errors per Seed (EPS)\\ \end{tabular}} & {Errors per Coverage (EPC)} & \multicolumn{1}{c|}{\begin{tabular}[c]{@{}c@{}}{Errors per Time (EPT)}\\ \end{tabular}} \\ \hline
Baseline 1 & 0\//5 = 0 &  0\//79.99 = 0  & 0\//300 = 0 \\ \hline
Baseline 2 & 0\//1 = 0 &  0\//78.06 = 0  & 0\//300 = 0 \\ \hline
Baseline 3  & 1\//19 = 0.052 &  1\//80.04 = 0.012  & 1 \//293 = 0.003 \\ \hline
{\bf{\tool}} &  {\bf 2\//12 = 0.167}  &  {\bf 2\//78.08 = 0.026}  & {\bf 2\//227 = 0.008} \\ \hline
\end{tabular}
\end{table}

In Fig.~\ref{fig:rq1_coverage} and Fig.~\ref{fig:rq1_corpus_size}, we
compare the actual coverage achieved with the ICSes from all
approaches and their corresponding sizes for all Java programs. While
a larger initial corpus or higher code coverage can enhance the
likelihood of detecting bugs, it is essential to balance these factors
with the effectiveness of bug detection. An excessively large corpus
may result in increased time and resource consumption during
subsequent fuzzing
phase~\cite{klees2018evaluating,rebert2014opt,skyfire,moonlight}. For
instance, consider program 12, in which Baseline 3 has a 483.3\%
larger ICS than \tool, but still covers less code than \tool (74.2\%
v/s 85\%). Such a trend is observed in {\em 27 of the 50} Java
programs, highlighting the importance of the quality of the ICSes to
potentially capture more bugs.
%

As shown in Table~\ref{tab:rq2-results}, {\tool} excels in all three
efficiency metrics.  For {\em \underline{Errors per Seed}}, Baseline~1 exhibits an
EPS of 0, as it fails to trigger any errors within the 5-minute time
limit.
This result can be attributed to the fact that none of the five
randomly generated seeds by JQF immediately trigger exceptions. A
similar result is observed with Baseline 2, where no errors are
detected in the initial corpus. This outcome is likely due to the
nature of the 50 randomly generated seeds, which fail to trigger
exceptions before the minimization process (thus, also fail after
that). That is, the randomly generated seeds in Baseline 2 do
not contain inputs capable of inducing errors.

In contrast, Baseline 3 triggers an average of one error per 19 seeds (or {\em $\sim$5 errors per 100 seeds})
after the fuzzing and minimization process. 
Compared to \tool, this is 221.2\% lower, which triggers the same number of errors per 6 seeds in the ICS (or {\em $\sim$16 errors per 100 seeds}).
This indicates that \tool is more efficient in producing the initial corpus, capturing more errors with a smaller size, as opposed to using JQF's intial corpus.
The superior performance of {\tool} can be attributed to the use of LLMs ({\em e.g.}, GPT) for a targeted seed generation, which results in generating error-triggering inputs, unlike the baselines, which rely on random generation.


In the analysis of the {\em \underline{Errors per Coverage}} (EPC)
metric (refer to Table ~\ref{tab:rq2-results}), the results reveal a
clear distinction between the performance of the baselines and
{\tool}. For Baseline 1 and Baseline 2, the EPC values are both zero
(0\//79.99 = 0 and 0\//78.06 = 0, respectively). This is consistent
with the fact that neither baseline was able to trigger any errors
during the evaluation, despite achieving substantial code
coverage. The failure to trigger errors can be attributed to the
randomly generated seeds, which, though providing good coverage of the
code under test (an average of almost 80\% of statement coverage),
lacked the specific inputs necessary to induce exceptions or errors.
Baseline 3 shows a marginal improvement with an EPC of 0.012, {\em
  i.e.}, on an average, triggered 1 error while achieving a code
coverage of 80.04\%.
However, this is {\bf 116.7\%} lower than \tool, which detected
$\sim$2 errors for every 78\% of covered code.
These results indicate that the seeds in the ICS produced by \tool not
only explore different execution paths, but are also effective in
uncovering diverse errors, {\em thereby achieving a better balance
  between code coverage and error detection.}


In the analysis of {\em \underline{Errors per Time}} (EPT) metric
(time-efficiency), both Baseline 1 and Baseline 2 attain an EPT of 0
(Table~\ref{tab:rq2-results}), {\em i.e.}, no errors were triggered
during the fixed time limit of 300 seconds (5 minutes).  This suggests
that the random generation of seeds by JQF is generally ineffective in
producing error-triggering inputs in the given timeframe.  Baseline 3
achieves an
EPT value of 0.003.
This suggests that Baseline 3, which utilizes a
combination of random seed generation and blind mutation followed by corpus
minimization, is relatively more efficient than Baselines 1 and
2. However, the low time-efficiency value indicates that error
detection occurs slowly, with only one error captured every 293 seconds ({\em i.e.}, the average time taken for the fuzzing process to conclude for all programs).
In contrast, {\tool} exhibits a higher time-efficiency of 0.008 (by 166.7\%),
indicating that {\em 2 errors were triggered within 226 seconds}, 
highlighting the improved efficiency of {\tool} in identifying bugs in
comparison to all baselines.  Unlike the random generation strategies
employed by the baselines, {\tool} focuses on maximizing code coverage
and error detection (see Section~\ref{sec:rq3}), resulting in faster bug
detection within the same fixed time window.

\subsubsection{Results for Python}

\begin{figure}[t]
\begin{center}
\includegraphics[width=3.4in]{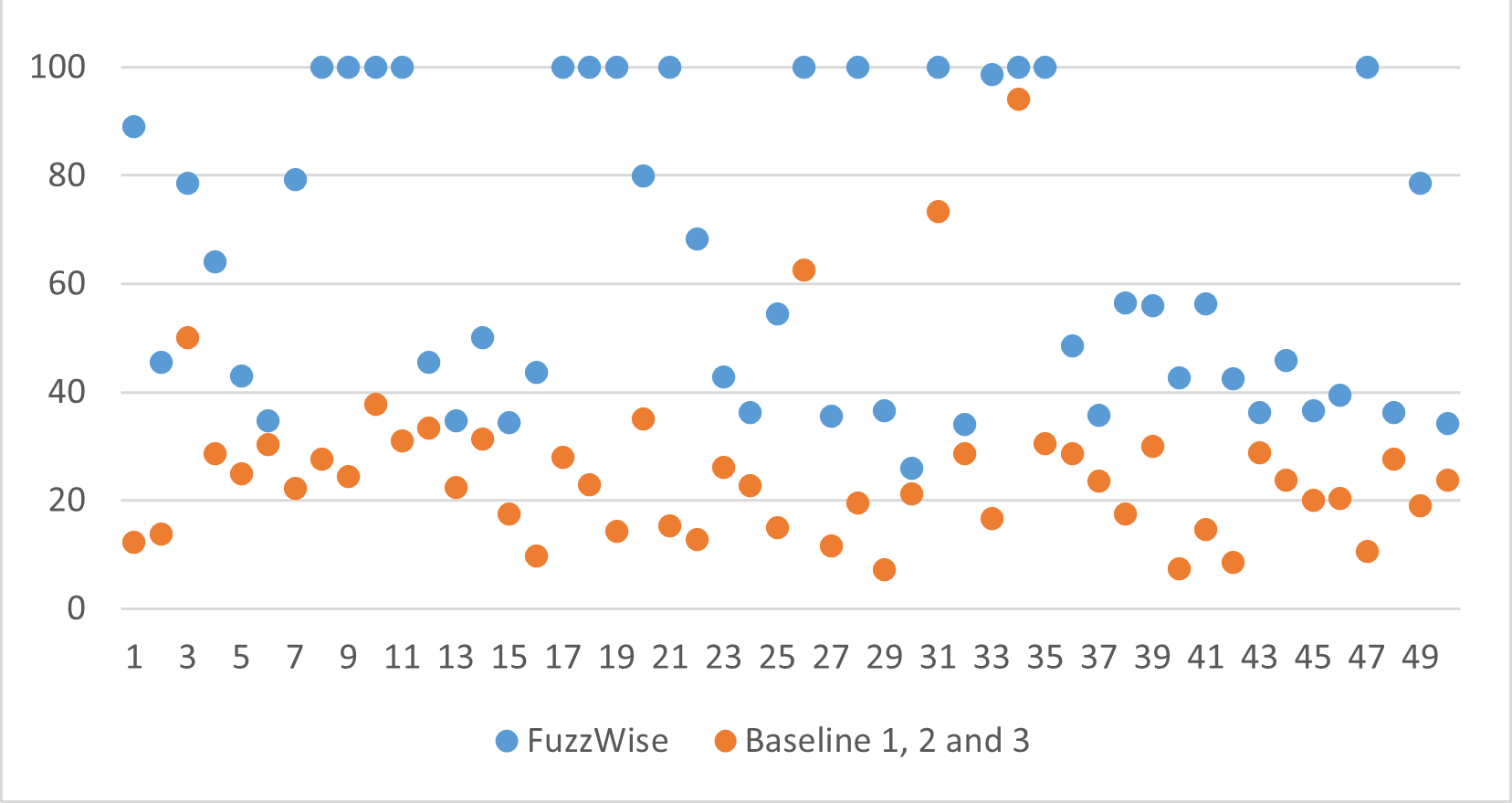}
\vspace{-9pt}
\caption{Coverages in All \underline{Python} Programs (RQ2)}
\label{fig:rq1_python_coverage}
\end{center}
\end{figure}

\begin{figure}[t]
\begin{center}
\includegraphics[width=3.4in]{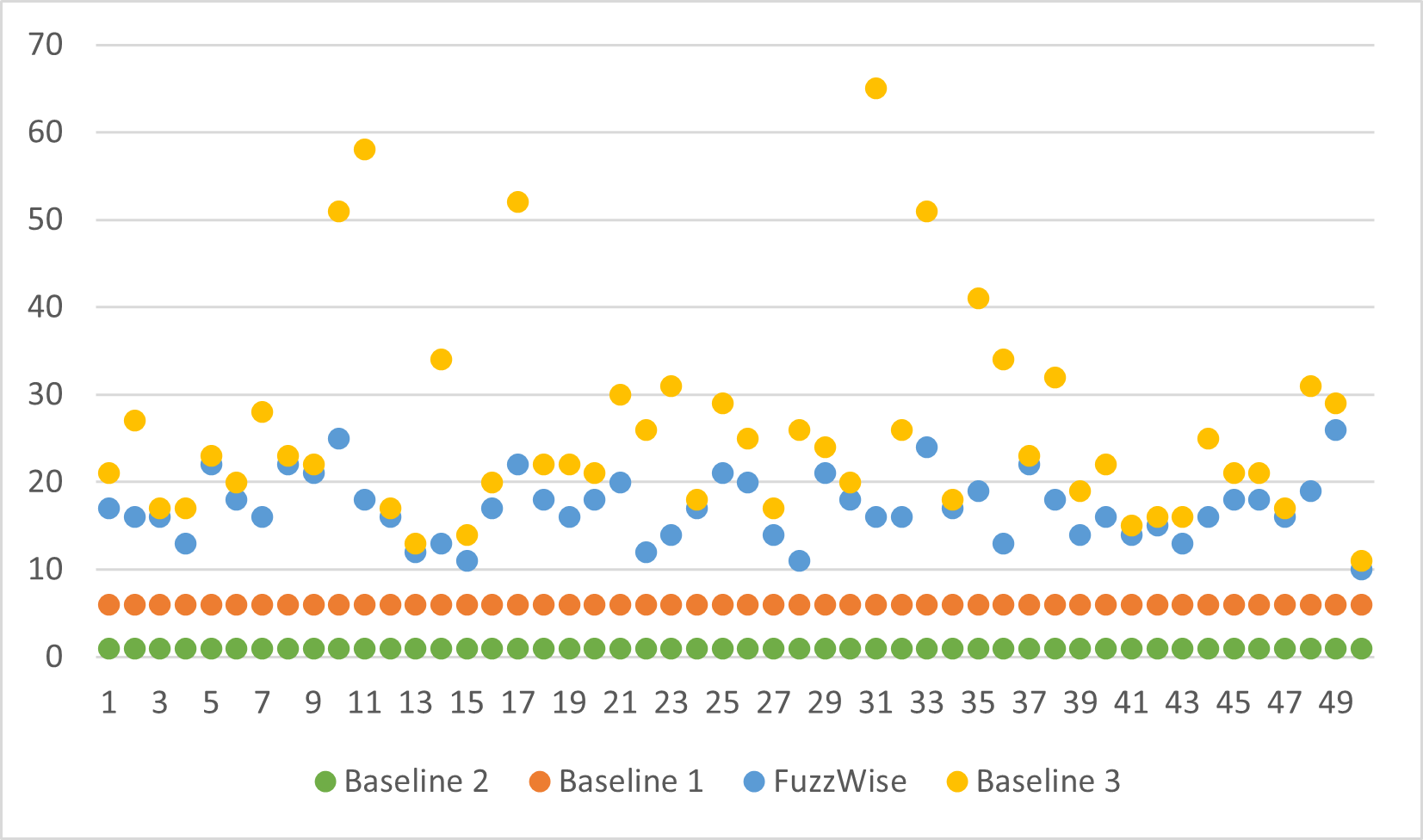}
\vspace{-9pt}
\caption{Corpus Sizes for All \underline{Python} Programs (RQ2)}
\label{fig:rq1_python_corpus_size}
\end{center}
\end{figure}

\begin{table}[t]
\small
\caption{Efficiency in Generating the Initial Corpus of Seeds for \underline{Python} (RQ2)}
\vspace{-6pt}
\label{tab:rq2-results-python}
\centering
\begin{tabular}{|l|c|c|c|}
\hline
& \multicolumn{1}{c|}{\begin{tabular}[c]{@{}c@{}}Errors per Seed (EPS)\\ \end{tabular}} & {Errors per Coverage (EPC)} & \multicolumn{1}{c|}{\begin{tabular}[c]{@{}c@{}}{Errors per Time (EPT)}\\ \end{tabular}} \\ \hline
Baseline 1 & 0\//5 = 0 &  0\//24.81 = 0  & 0\//300 = 0 \\ \hline
Baseline 2 & 0\//1 = 0 &  0\//24.81 = 0  & 0\//300 = 0 \\ \hline
Baseline 3  & 1\//19 = 0.052 &  1\//25 = 0.04  & 1\//124 = 0.008 \\ \hline
{\bf{\tool}} &  {\bf 3\//11 = 0.27}  &  {\bf 3\//64.89 = 0.04}  & {\bf 3\//183 = 0.016} \\ \hline
\end{tabular}
\end{table}

As illustrated in Fig.~\ref{fig:rq1_python_coverage} and Fig.~\ref{fig:rq1_python_corpus_size}, the actual code coverage and corresponding initial corpus sizes for Python programs exhibit similar trends as observed for Java ones (see Section~\ref{sec:rq1-results}).
We can see that, on an average, {\tool} (depicted in blue in both
figures) tends to produce {\em smaller corpora with higher code coverages}.
Overall, as seen in Table ~\ref{tab:rq2-results-python}, {\tool}
produces competitive results along all dimensions of initial corpus quality.

For {\em \underline{Errors per Seed} (EPS)}, baselines 1 and 2 (each having an
EPS of 0) failed to trigger any errors with their 5 and 50 randomly
generated seeds, respectively. 
This can be attributed to the randomness in Py-AFL's seed generation
and Py-AFL-cmin's minimization processes, respectively.  Baseline~3
shows a slight improvement, triggering one error per 19 seeds
generated (or {\em $\sim$5 errors per 100 seeds}), giving an EPS of
0.052. While this indicates that some error-triggering seeds were
generated, the performance is limited by the random mutating
nature of Py-AFL, which is less effective than our LLM-based prompting
strategy to target vulnerable areas in code.


In contrast, {\tool} shows a higher EPS of $3/11 = 0.27$,
demonstrating that, on an average, one error is triggered for every
$\sim$4 seeds (or {\em $\sim$27 errors per 100 seeds}). This is over
{\bf 5$\times$} more efficient than Baseline 3. The improvement can be
attributed to {\tool}'s use of LLMs to generate targeted seeds,
allowing it to generate seeds that are specifically designed to
increase the likelihood of triggering errors, rather than relying on
random generation and blind mutation strategies.


Regarding {\em \underline{Errors per Coverage}} (EPC), Baselines 1 and
2 once again did not trigger any errors, with each covering only
24.81\% of the code on average. The inability to detect any bugs
highlights the limitations of random seed generation. It also
underscores the fact that high code coverage alone is not enough for
effective bug detection, as randomly generated seeds often fail to
target vulnerable areas of the code. Baseline 3 shows a modest EPC of
0.04, as it covers 25\% of the code and triggers one error. While this
is a slight improvement over the other baselines, the fuzzing process
still lacks precision, resulting in only minimal error detection
relative to the coverage achieved. Compared to the average coverage
achieved by JQF-AFL for Java (80.04\% in Table~\ref{tab:rq2-results})
on the same average initial corpus size ({\em i.e.}, 19), the average
coverage achieved by Py-AFL for Python is significantly lower (by
220.2\%). This indicates the relatively weaker fuzzing capabilities
of Py-AFL.

{\tool} also achieves an EPC of 0.04, but this must be
interpreted in the context of its higher number of detected bugs (3
bugs) and overall coverage of 62.28\%. The dual prompting strategy
employed by {\tool}, which initially focuses on maximizing coverage
and later targets error-prone regions, results in a broader
exploration with the discovery of more paths in a program. The wider coverage obtained by \tool
allows it to explore the test case space effectively and find diverse errors in more areas, highlighting the overall
benefits of its design.

In the case of time efficiency, Baselines 1 and 2 each have an EPT of
0. Despite 300 seconds of execution, no errors were found, reaffirming
the inefficiency of the baselines' strategies, where neither random
seed generation nor the \code{AFL-cmin} minimization is effective in
triggering errors. Baseline 3 triggers an average of one error
over a time period of 124 seconds, resulting in an EPT of
0.008. This very low rate of error detection shows the
inefficiency of Py-AFL's fuzzing when~it comes to generating
error-triggering seeds within the time limit. The random nature
of seed generation combined with the time-consuming
minimization limits the baseline's error-finding capabilities.

{\tool} outperforms all baselines, with an EPT of 0.016, detecting two
errors in 178 seconds, which is over {\bf 2$\times$} better than
Baseline 3. This superior performance is due to a combination of
{\tool}'s use of LLM for targeted seed generation and its dual
prompting strategy. The LLM enables the generation of seeds
specifically aimed at either coverage or error detection, avoiding the
inefficiencies of random seed generation. Moreover, the dual prompting
strategy (see Section~\ref{sec:rq3}), which focuses first on
maximizing coverage and then shifting to error detection, optimizes
time efficiency, allowing more errors to be detected more quickly than
the baseline approaches.

\section{Evaluation on Prompting Strategy to Detect Errors (RQ3)}
\label{sec:rq3}

\subsection{Experimental Methodology}


{\tool} uses a combination of LLM-based Test Case Generation (TCG LLM)
and LLM-based Code Coverage Prediction ({\em i.e.},
CodePilot~\cite{forge24}). The performance of CodePilot has already
been documented in its original paper~\cite{forge24}.  In this
experiment, we focus on assessing the effectiveness of our prompting
strategy in TCG LLM, where we employ two distinct prompts: one
designed to guide the LLM to maximize code coverage and another aimed
at triggering or discovering errors, if present. {\tool} initially
generates test cases aimed at increasing code coverage. Once the
predicted coverage reached 100\%, the focus is then shifted to
generating test cases designed to trigger errors. To evaluate the
significance of this prompting strategy, we compared {\tool} with a
variant where the TCG LLM uses a single prompt focused only on
maximizing coverage of the PUT, similar to a coverage-guided fuzzer,
throughout the entire time limit.
Both variants were run for a fixed duration of 5 minutes on the
Python dataset, measuring the same metrics as in RQ1.

\subsection{Empirical Results}
\label{sec:rq3-results}

\begin{figure}[h]
\begin{center}
\includegraphics[width=3.4in]{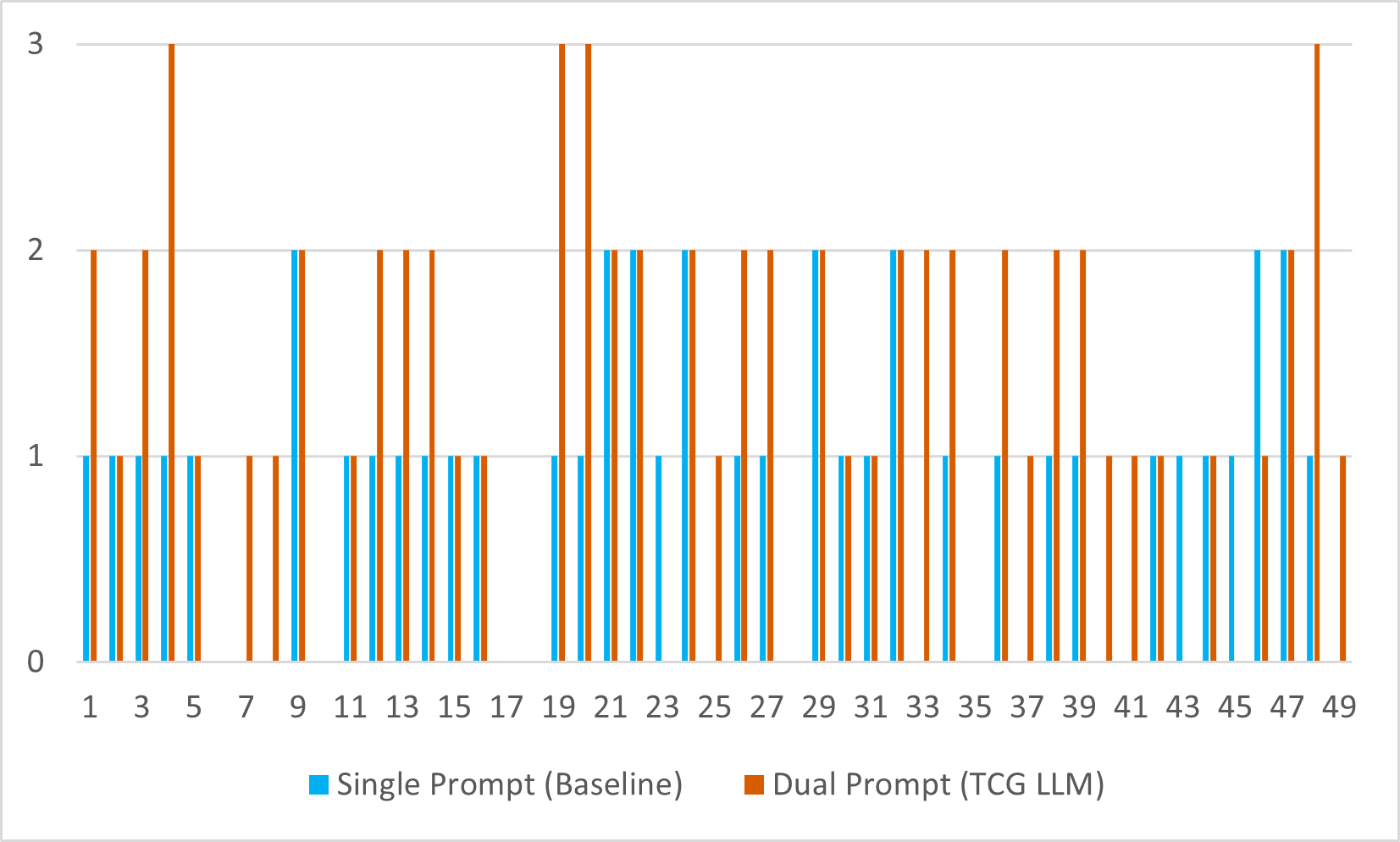}
\vspace{-9pt}
\caption{Number of Unique Detected Errors for Python Programs for Single and Dual-Prompt Settings (RQ3)}
\label{fig:rq3_errors}
\end{center}
\end{figure}

Fig.~\ref{fig:rq3_errors} illustrates the performance of the dual
prompting mechanism employed by {\tool} compared to a single prompting
mechanism. {\tool} significantly outperforms the baseline-which uses a
single prompt aiming to increase only code coverage. Specifically,
seeds generated by TCG LLM in {\tool} with dual-prompts identify an
equal or greater number of errors in 40 out of 50 ({\em i.e.}, {\bf 80\%}) programs than the baseline with a single prompt (aiming for
code coverage only). This improved performance can be attributed to the
dual prompt's ability to optimize seed generation by focusing first on
increasing code coverage and then targeting error detection.

In addition, in 6 out of 50 ({\em i.e.}, 12\%) programs, neither
{\tool} nor the baseline identifies any errors. 
When initialized to run for 5 minutes, the traditional fuzzer ({\em i.e.}, {\em Py-AFL}) fails to capture any errors for these 6 programs as well.
This lack of error
detection in these samples is likely due to the inherent complexity of the code snippets, which may require more sophisticated or
extended fuzzing time. In 4 out of 50 ({\em i.e.}, 8\%) code snippets,
the baseline triggers one more error than {\tool}. 
This may be attributed to the baseline’s single prompting strategy having a slight advantage in specific scenarios where the error-triggering potential of randomly generated seeds happens to align with particular conditions in those snippets. However, this does not diminish the effectiveness of our strategy, which targets error detection more deliberately. {\tool}'s dual prompting strategy is designed to comprehensively maximize both coverage and error detection. Therefore, even if the baseline performs marginally better in a few cases, {\tool} ensures more consistent and effective error detection across the entire dataset.

In summary, {\em our dual-prompt strategy, which targets both error
detection and coverage improvement, outperforms the single-prompt
strategy focused solely on increasing code coverage}. 
As outlined in Key Idea 1 (see Section~\ref{sec:key-ideas}), these findings further reinforce the benefits of a unified process of constructing the ICS in \tool,
where test-case generation can be dynamically adjusted. This result
explains why {\tool} performs better in previous
experiments than Baselines 1, 2, and 3, which primarily depend
only on seed generation to enhance only code coverage.

\section{Usefulness of {\tool} in Fuzzing (RQ4)}
\label{sec:rq4}

\subsection{Experimental Methodology}

\subsubsection{Procedure and Baseline}

In this experiment, we aim to evaluate {\tool}'s usefulness in the
fuzzing process. To achieve that, we assess the impact of the initial
corpus generated by {\tool} in the bug detection performance of
JQF-AFL~\cite{JQF}, a fuzzer for Java code. Specifically, we first use
{\tool} to produce an intial corpus for each Java program in our
dataset. The generated ICSes are then plugged into JQF-AFL, which uses
its built-in fuzzing mechanism to mutate and explore new execution
paths in the programs for a fixed time limit of 5 minutes. For the baseline, we chose the JQF-AFL fuzzer and its
default algorithm to produce the initial corpora through random
generation and mutation~\cite{JQF}. Upon completion, we compared the results of
running JQF-AFL by itself and JQF-AFL with the ICSes generated from
{\tool} (let us call it {\tool}+AFL).

\subsubsection{Metrics}
To evaluate the quality of {\tool}'s and the baseline's initial
corpora, we use the following metrics: 1) the average number of detected
errors after the time limit for the entire fuzzing process, and 2) the average percentage of code covered in 5 minutes of fuzzing, showing how well the initial corpus
helps the fuzzer explore different execution paths.




\subsection{Empirical Results}
\label{sec:rq4-results}

\begin{table}[t]
\small
\caption{Usefulness of Initial Corpus by {\tool} in Fuzzing (RQ4)}
\vspace{-6pt}
\label{tab:rq4-results}
\centering
\begin{tabular}{|l|c|c|}
\hline
& \multicolumn{1}{c|}{\begin{tabular}[c]{@{}c@{}}Avg. \#-of Detected Errors \\ \end{tabular}} & \multicolumn{1}{c|}{\begin{tabular}[c]{@{}c@{}}{Avg. \% of Statements Covered}\\ \end{tabular}} \\ \hline
JQF-AFL (Baseline) & 1  & 80.45\% \\ \hline
{\bf{\tool}+AFL} &  {\bf 2}    & {\bf 89.35\%} \\ \hline
\end{tabular}
\end{table}

In comparing the performance between JQF-AFL (the baseline) and
{\tool}+AFL, two key metrics—illustrated in Table
~\ref{tab:rq4-results}—highlight the clear advantage of {\tool}. JQF-AFL
alone triggers an average of one error per program in 5 minutes of fuzzing, whereas
{\tool}+AFL detects 2 errors. This result indicates that the initial
corpora from {\tool} contains better seeds than JQF-AFL, enabling the
fuzzing process to be more effective in detecting runtime errors.

Furthermore, {\tool}+AFL achieves an average code coverage of {\bf 89.35\%}
for a program in our dataset, compared to 80.45\% for the
baseline. This 11\% relative improvement reflects {\tool}’s ability to
explore more uncovered code paths through its coverage-maximizing
prompts in the same time frame, in contrast to AFL’s random test case
generation. This metric also explains the reason for the higher number
of bugs detected by {\tool} in Table~\ref{tab:rq4-results}.



\begin{wrapfigure}{r}{2.7in}
\small
	\centering
	\lstset{
		numbers=left,
		numberstyle= \tiny,
		keywordstyle= \color{blue!70},
		commentstyle= \color{red!50!green!50!blue!50},
		frame=shadowbox,
		rulesepcolor= \color{red!20!green!20!blue!20} ,
		xleftmargin=1.5em,xrightmargin=0em, aboveskip=1em,
		framexleftmargin=1.5em,
                numbersep= 5pt,
		language=Java,
    basicstyle=\scriptsize\ttfamily,
    numberstyle=\scriptsize\ttfamily,
    emphstyle=\bfseries,
                moredelim=**[is][\color{red}]{@}{@},
		escapeinside= {(*@}{@*)}
	}
\begin{lstlisting}[]
public class Main {
  static int N = 0, M = 0, R = 0;      
  static int[][] d;
  static int[] r;
  static int A = 0, B = 0, C = 0;      
  static int res = 0;
  static boolean used [];
  public static void main(String[] args){
    Scanner sc = new Scanner(System.in);
    N = sc.nextInt();
    M = sc.nextInt();
    R = sc.nextInt();
    d = new int[201][201];
    r = new int[9];
    used = new boolean[9];
    for(int i = 1; i <= N; i++) {      
      for(int j = 1; j <= N; j++) {    
        if (i != j) {
          d[i][j] = Integer.MAX_VALUE; 
        } } }
    for(int i = 1; i <= R; i++) {      
      r[i] = sc.nextInt(); }
    for(int i = 1; i <= M; i++) {      
      A = sc.nextInt();
      B = sc.nextInt();
      C = sc.nextInt();
      if(d[A][B] > C) {
        d[A][B] = d[B][A] = C; } }
    for(int k = 1; k <= N; k++) {      
      for(int i = 1; i <= N; i++) {    
        for(int j = 1; j <= N; j++) {  
          if(d[i][j] > d[i][k] + d[k][j]) {
            d[i][j] = d[i][k] + d[k][j];
          } } } }
    res = Integer.MAX_VALUE;
    dfs(1, -1, 0);
    System.out.println(res);
  }
  private static void dfs(int c, int las, int dist) {
    if(c == R + 1) {
      if(res > dist) {
        res = dist; }
      return; }
    for(int i = 1; i <= R; i++) {      
      if(!used[i]){
        used[i] = true;
        if(las == -1) dfs(c+1, i, 0);  
        else dfs(c+1, i, dist+d[r[las]][r[i]]);
        used[i] = false;
      } } } }
\end{lstlisting}
\vspace{-16pt}
\caption{An Example of a Java Program for Fuzzing}
\label{fig:rq4-example}
\end{wrapfigure}

Next, let us consider the example in
Fig.~\ref{fig:rq4-example} for further analysis. The comparison between JQF-AFL's fuzzing
and the integration of itself with {\tool}'s initial corpus
demonstrates the impact of seed quality in ICS on the effectiveness
and efficiency of the fuzzing process.

For this Java program, with JQF-AFL's initial corpus, the fuzzer
JQF-AFL achieved 100\% code coverage within 5 minutes, processing a
full corpus size of 58 seeds. However, it {\em failed to detect any
  errors during this time frame}.  This demonstrates the lower-quality
of the initial corpus generated by JQF-AFL through random generation
and mutation.  While exploring more code structure, JQF-AFL does not
generate effective test inputs that trigger errors. The generated test
cases from JQF-AFL's initial corpus are effective in covering the
statements, but they may lack the specificity or targeted mutations
necessary to generate inputs that trigger errors.


On the other hand, {\em when using the initial seed corpus generated
  by {\tool}, JQF-AFL also achieves 100\% code coverage in the same
  time frame. However, a key difference is that {\tool}+AFL detects a
  total of {\bf 8 errors}}, a significant improvement in effectiveness
when compared with the fuzzing process with the ICS from JQF-AFL.
Moreover, this was achieved with {\em a smaller initial corpus of 39
  seeds ({\em i.e.}, 48.7\% lower)}, showing that fewer, but more
targeted, test cases are more effective in bug detection and in
achieving relatively better results.  This result shows that {\em the
quality of the seeds in the initial corpus created using {\tool} is
higher, meaning fewer seeds are needed to explore almost the same
amount of code and trigger more errors}. This smaller, more focused
corpus reduces the computational overhead and fuzzing time required,
enabling faster and more efficient bug detection.


While achieving full code coverage is important, it is not always
sufficient for detecting bugs. {\tool} enhances fuzzing by generating
a more error-targeted seed corpus with high coverages. By combining
targeted test case generation with the fuzzing power of JQF-AFL,
{\tool} demonstrates how a smaller, more refined corpus can outperform
a larger, randomly generated one. This shows the importance of seed
quality in driving the effectiveness and efficiency of the fuzzing.

\section{Limitations and Threats to Validity}
\label{sec:limitations}

{\em Limitations}. First, while we can successfully fuzz programs of
certain size (150+ lines of code), we aim to enhance scalability in
future work. Current limit is due to the underlying LLM. The latest
models such as Claude~\cite{claude-ai} can handle a large size of
input code sequence (200k).

Second, due to their complex nature and reliance on probabilistic algorithms, black box LLMs may exhibit unpredictable behavior in certain outlying scenarios. Nevertheless, executing the generated test cases ensures their effectiveness in runtime error detection.  Third, although coverage prediction provides a reasonable estimation, it does not work as well for unseen libraries.

{\em Threats to Validity}. {First}, we do not perform a direct comparison with other fuzzers for both Java and Python. This is because collecting the ICSes as required for baseline comparison is not straightforward. Furthermore, many popular fuzzers like {\em Atheris}~\cite{atheris} and {\em Pythonfuzz}~\cite{pythonfuzz} lack corpus distillation techniques, making them unsuitable as appropriate baselines for our evaluation. {Second}, in the computation of Errors per Time (EPT) metric used in RQ2 (Section~\ref{sec:rq2}), \tool uses GPT as the LLM and records its corresponding inference times. However, these results could vary based on the latency associated with the API calls for GPT or other LLMs, or also vary based on the hardware used while inferring with open-weight LLMs like Llama~\cite{code_llama}.

\section{Implications}

{\em Novelty.} In this work, we show that a tandem of LLMs in test case generation and code coverage prediction could operate well in constructing
a high-quality initial corpus of seeds for the fuzz testing
process. An interesting finding is that guiding the LLM in generating
test cases in targeting bug detection as well as improving code
coverage is more effective than guiding it in improving code coverage
only, especially when we aim to have a minimal size of the initial
corpus of seeds.

{\em Implications.} The findings of this research open up promising
directions for future work in several key areas where actual execution
of test cases is undesirable, infeasible, or inefficient:

1. {\em Bug Detection in Incomplete Code}: The ability of {\tool} to predict code coverage contributions without requiring actual execution suggests potential for applications in bug detection for incomplete or under-development code. Traditional fuzzing approaches require complete, executable code to test. However, {\tool} could be adapted to work in scenarios where the code is not fully functional, allowing developers to identify potential vulnerabilities or areas of interest before the software is finalized. Future work could explore leveraging {\tool} for early-stage bug detection, accelerating the development by identifying issues earlier in the software lifecycle.

2. {\em Test Case Prioritization Without Execution}: In environments where running test cases to determine their prioritization is resource-intensive or impractical, {\tool}'s predictive model offers a way to prioritize test cases based on their expected contribution to code coverage. This could be extended to prioritize tests not only for coverage but also for other quality metrics like bug likelihood or criticality, without the need to execute them. Future research could focus on refining and expanding predictive models to handle diverse prioritization strategies, particularly in continuous integration pipelines where time and resources are often constrained.

3. {\em Minimization of Test Corpus Without Execution}: The integration of corpus generation and minimization in {\tool}, bypassing the need for actual execution, presents an efficient way to optimize the size and relevance of test suites. This approach could be extended to scenarios where minimizing large test suites is critical but execution is costly or impractical. For instance, cloud-based environments or systems with extensive setup requirements could benefit from a predictive, non-execution-based approach to test suite minimization.

4. {\em Scenarios with Complex or Impossible Execution}: {\tool}'s ability to function without execution is particularly relevant in scenarios where execution is impossible due to missing dependencies, lack of access to a target system, or high complexity in setup. For instance, in embedded systems, IoT devices, or proprietary environments where execution might be restricted or costly, the use of predictive models could help generate meaningful tests or corpus subsets that maximize coverage. Future studies could adapt {\tool} to work in these specialized environments, providing a cost-effective alternative to traditional fuzzing.


\section{Related Work}
\label{sec:related}

Fuzzing has been extensively studied in software
testing. Miller {\em et al.}~\cite{miller1995fuzz} conducted an empirical
study evaluating the reliability of UNIX utilities, pioneering the
concept of fuzzing by exploring the effectiveness of providing
invalid, unexpected, or random data as input to uncover bugs.


As fuzzing progresses, efforts shifted to enhancing the process's
efficiency and effectiveness. {\em Coverage-guided
  fuzzing}~\cite{10.1145/3293882.3339002,JQF,10.1145/3133956.3138820}
emerged as an advancement, addressing the need for a systematic
approach to explore the vast input space of software. By monitoring
code execution with each input and prioritizing inputs exploring new
paths, coverage-guided fuzzers like AFL~\cite{JQF} discover
deep-seated bugs and vulnerabilities. Bohme {\em et
  al.}~\cite{10.1145/3133956.3134020} introduced coverage-based
greybox fuzzing as a Markov Chain, presenting AFLFast as an extension
of AFL that significantly increases path coverage. 
GreyOne~\cite{244046} is a coverage-guided greybox fuzzer that
incorporates data flow analysis to prioritize paths that are more
likely to lead to vulnerabilities. Pasareanu and
Visser~\cite{10.1145/3364452.3364455} surveyed new trends in symbolic
execution relevant to fuzz testing. Cadar {\em et
  al.}~\cite{10.5555/1855741.1855756} introduced KLEE, a symbolic
virtual machine built on top of LLVM that uses symbolic execution to
systematically explore various paths.



Recent advancements in fuzzing techniques have shown promise in
overcoming the coverage plateau issue.
CODAMOSA~\cite{10.1109/ICSE48619.2023.00085} leverages the synergy
between search-based software testing (SBST) and LLMs to push beyond
the coverage plateau. By integrating SBST with LLMs, CODAMOSA explores
the input space via the embeddings of input values to generate more
diverse and sophisticated unit tests. Fuzz4All~\cite{xia2024fuzz4all}
can work across multiple programming languages, combined with its
uses of LLMs for input generation.



\section{Conclusion and Future Work}
\label{sec:conclusion}

The proposed coverage-guided fuzz testing framework, named {\tool},
addresses challenges faced by existing frameworks by leveraging a LLM
for code coverage prediction and test case generation. It prioritizes
high-coverage test cases and employs a feedback loop to refine test
cases.
Additionally, instead of traditional test mutation techniques, we
employ the LLM to automatically generate test cases. These test cases
undergo a feedback loop, where those contributing to higher code
coverage are retained, while others are reintroduced to the LLM for
refinement.
Empirical evaluation demonstrates {\tool}'s superiority in
detecting runtime errors/exceptions and its potential to enhance
defect detection.

{\bf Future Work.} we aim to expand the capabilities of {\tool} in
key areas. Firstly, we will explore incorporating other
programming languages to enhance its applicability across diverse
software projects. We also intend to refine the feedback loop
mechanism by integrating more sophisticated models to better identify
and retain high-coverage test cases. Another area of interest is
optimizing the computational efficiency of {\tool} further, making it
more scalable.

Lastly, we will investigate the possibility of integrating {\tool} with other automated testing tools and frameworks to create a more holistic and seamless testing environment. This would include enhancing the detection of more complex runtime errors and exceptions.


\section{Data Availablity}

Our code and data is available in our project's
website~\cite{fuzzwise-website}.


\newpage

\balance

\bibliographystyle{ACM-Reference-Format}

\bibliography{references,references-planning}

\end{document}